\documentclass[12pt,letterpaper]{article}
\usepackage{amsfonts}
\usepackage{amssymb}
\usepackage{graphicx}
\usepackage{osajnl}

\DeclareRobustCommand{\baselinestretch{2}}

\begin{document}

\title{Resonant interaction of optical pulses with plasmonic oscillations in
metal nanoparticles}
\author{Ildar R. Gabitov$^{1,2}$, Robert Indik$^1$,
Natalia M. Litchinitser$^3$, Andrei I. Maimistov$^4$,
Vladimir M. Shalaev$^5$, Joshua E. Soneson$^6$}
\affiliation{$^1$: Mathematics Department, University of Arizona, 617 N. Santa Rita
Avenue, Tucson, AZ 85721 USA \\
$^2$: Theoretical Division, Los Alamos National Laboratory, Los Alamos, NM
87545 USA \\
$^3$: Department of Electrical Engineering and Computer Science,
University of Michigan, 525 East University
Avenue, Ann Arbor, Michigan 48109 USA \\
$^4$: Department of Solid State Physics, Moscow Engineering Physics
Institute, Kashirskoe sh. 31, Moscow, 115409 Russia \\
$^5$: School of Electrical and Computer Engineering, Purdue University, West
Lafayette, IN 47907 USA \\
$^6$: Program in Applied Mathematics, University of Arizona, 617 N. Santa
Rita Avenue, Tucson, AZ 85721 USA}

\begin{abstract}
We derive envelope equations which generate the Maxwell-Lorentz model and
describe the interaction of optical pulses
with plasmonic oscillations in metal nanoparticle composites. A
family of solitary wave solutions is found which is analogous to self-induced
transparency in Maxwell-Bloch. The evolution of incident optical pulses
is studied
numerically as are the collision dynamics of the solitary waves.
These simulations reveal that the collision dynamics vary from near
perfectly elastic to highly radiative depending on
the relative phase of the initial pulses.
\end{abstract}


\maketitle

Quantum effects in metal nanoparticles driven by a resonant optical field
play an important role in inducing a strong nonlinear response,
as was recently shown~\cite{Rautian,Drachev}.
In this Letter we consider the nonlinear resonant interaction
of ultrashort optical pulses with metal nanoparticles distributed
uniformly in a host medium. We restrict to the case of composite
materials for which the resonance frequencies of the host medium are
well separated from those of the nanoparticles. Examples include
silver or gold spherical or spheroidal nanoparticles embedded in SiO$_2$.
In these cases, the plasmonic resonance frequencies are in the visible
part of the spectrum
while the resonance of the host is in the ultraviolet. 

Light interaction with metal nanoparticles can be described by a
system consisting of Maxwell's equations for the electric field, and an
oscillator equation describing the displacement of conduction electrons
in the metal nanoparticles from equilibrium (plasmonic oscillations).
The nanoparticles are much
smaller than the optical carrier wavelength $\lambda _{0}$. This allows
light scattering and spatial effects in the nanoparticles to be neglected.
As shown by Rautian \cite{Rautian} and Drachev, et al. \cite{Drachev}, who
further developed the earlier work by Hache, et al. \cite{Hache},
the response of the conduction electrons in the metal nanoparticles to an
external electric field induces a leading-order cubic nonlinearity.
The interaction of the electric field with plasmonic
oscillations in nanoparticles with resonance frequency $\omega_{r}$ in the
presence of this cubic nonlinearity can be described by the forced
Duffing equation
\begin{eqnarray}
\tilde{\mathcal{Q}}_{TT}+\omega _{r}^{2}\tilde{\mathcal{Q}}+\varkappa \tilde{
\mathcal{Q}}^{3}=(e/m)\tilde{\mathcal{E}}.
\label{duffing}
\end{eqnarray}
In this expression $\tilde{\mathcal{Q}}$ represents plasmon
displacement from equilibrium, $T$ is time, $\varkappa$ is the
coefficient of nonlinearity, $e$ and $m$ are the electron charge
and rest mass, respectively, and $\tilde{\mathcal{E}}$ is the
electric field. The tilde is used to denote rapidly-varying
quantities. The nonlinear coefficient $\varkappa$ can be
estimated by comparing the off-resonance nonlinear response in
Eq. (\ref{duffing}) with the Drude nonlinearity for
(non-resonant) conduction electrons in metal nanoparticles. That
susceptibility is characterized by \cite{Rautian} $\chi
^{(3)}\simeq Ne^{4}a^{2}/(m\hbar ^{2}\omega_0 ^{4})$. Here $a$
and $N$ are the radius of the nanoparticle and the conduction
electron density of the metal, respectively, and $\omega_0$ is the optical
carrier frequency. This results in the estimate
$\varkappa\simeq(ma\omega_0 ^{2}/\hbar )^{2}$.

We are interested in pulse dynamics which vary on a much slower scale
than the plasmonic, host atom, and carrier wave oscillations, and can
be described using a slowly-varying envelope approximation. In this
approximation, Eq. (\ref{duffing}) becomes
\begin{eqnarray}
i\mathcal{Q}_{T}+(\omega_{r}-\omega_0)\mathcal{Q}+(3\varkappa /2\omega_0)|
\mathcal{Q}|^{2}\mathcal{Q}=-(e/2m\omega_0)\mathcal{E},
\label{Qfield}
\end{eqnarray}
where the slowly-varying envelopes of the electric field and plasmonic
oscillations are represented by $\mathcal{E}$ and $\mathcal{Q}$
respectively. Maxwell's equation couples to the material
polarization induced by the plasmonic oscillations. The equation for the
electric field envelope is
\begin{eqnarray}
i\left( \mathcal{E}_{Z}+\frac{1}{v_{g}}\mathcal{E}_{T}\right) =-\frac{2\pi
\omega_0N_{p}e}{cn_0}\left\langle \mathcal{Q}\right\rangle -\frac{2\pi \omega_0
N_{a}|d|^{2}}{cn_0\hbar \Delta_a}\mathcal{E}-\frac{2\pi i\omega_0 N_{a}|d|^{2}}{%
cn_0\hbar \Delta_a^{2}}\mathcal{E}_{T},
\label{maxwell}
\end{eqnarray}
where $Z$ is the propagation coordinate, $v_{g}$ is group velocity, $c$ is
the speed of light, $n_0$ is the refractive index evaluated at the carrier
frequency $\omega_0$, $N_{p}$ is the product of the conduction electron
density $N$ and the metal filling factor $p$ (the fraction of the composite
occupied by metal). $N_{a}$ is the concentration of host atoms, $d$ is
the projection of the dipole matrix
element in the direction of the electric field polarization, and $\Delta_a
=\omega_a-\omega_0$ is detuning from the resonance frequency of host
atoms. The last two terms in Eq.~(\ref{maxwell}) represent
corrections to the refractive index and group index due to the off-resonance
interaction with the host medium, which, for illustration, is considered as an
ensemble of two-level atoms. This equation is derived from the Maxwell-Bloch
equations in the non-resonant case by considering $\Delta_a$ as a large
parameter and applying the adiabatic following approximation~\cite{Basharov}.
Additional resonances would produce similar terms.
We consider the case where optical pulse intensity and duration as well as
composite
material parameters are such that the characteristic length of resonant light
interaction with plasmonic oscillations is much smaller than the
characteristic lengths for both group velocity dispersion and
nonlinearity induced by the host medium. Therefore the terms representing
these effects are omitted from Eq.~(\ref{maxwell}).

In a composite material, the sizes and shapes of metal
nanoparticles vary due to limited fabrication tolerances. It is
known that the plasmon resonance in spherical metal nanoparticles
depends weakly on size in the range between 10 and 50nm
\cite{Shalaev}, so that variations in size are not important.
However, variations in the shape and orientation of the
nanoparticles can significantly change plasmonic resonance
frequencies. This results in a broadening of the resonance line
of the bulk composite. The angle brackets $\left\langle
\mathcal{Q}(t,z,\omega )\right\rangle =\int_{-\infty }^{\infty
}\mathcal{Q}(t,z,\omega)g(\omega) \mbox{d}\omega$ denote
averaging over the distribution $g(\omega)$ of the resonance
frequencies (line shape). Defining
\begin{eqnarray}
\mathcal{E}=-E\frac{2 m
\omega_{0}^{3}}{e}\sqrt{\frac{2}{3\varkappa}}\exp{(ik_{s}Z)},~~~\mathcal{Q}=Q
\omega_{0}\sqrt{\frac{2}{3\varkappa}}\exp{(ik_{s}Z)},
\label{rescaling}
\end{eqnarray}
where $k_{s}=2\pi \omega_{0} N_{a}|d|^{2}/c n_{0}\hbar \Delta_a$ ,
and introducing the copropagating coordinate system
$z=(\omega_{p}^{2}/4cn_{0}\omega_{0})Z,~~~t=\omega_{0}(T-Z/u)$,
($u$ here is shifted group velocity defined as
$u^{-1}=v_{g}^{-1}+(2 \pi \omega_{0} N_{a}|d|^{2}/c n_{0}\hbar
\Delta_a^{2})$, $\omega_{p}^2=4\pi N_pe^2/m$,
$\omega=(\omega_r-\omega_0)/\omega_0$,
Eqs.~(\ref{Qfield}) and~(\ref{maxwell})  can be reduced to the
simpler form
\begin{eqnarray} iE_{z}=\left\langle
Q\right\rangle,\hspace{5mm} iQ_{t}+\omega Q+|Q|^{2}Q=E.
\label{NormalizedEq}
\end{eqnarray}
These equations represent a generalization of the classical
Maxwell-Lorentz model. In the case of identical nanoparticles,
the averaging in~(\ref{NormalizedEq})
 can be reduced to a single dimensionless frequency $\bar{\omega}$ [i.e.
detuning frequency distribution $g(\omega )=\delta (\omega
-\bar{\omega})$]. Under this condition the system has solitary
wave solutions:
\begin{eqnarray}
E(t,z)=\frac{v^{3/4}\exp \left[ i\varphi +i\Omega t-iK\xi -i\chi (\xi )%
\right] }{\xi _{0}[\cosh (\xi /\xi _{0})+K]^{1/2}},\,\,Q(t,z)=E(t,z)\frac{%
\exp \left[ -2i\chi (\xi )\right] }{\sqrt{v}}, \label{solutions}
\end{eqnarray}
where $\xi =[z-v(t-\tau )]/\sqrt{v}$, $\chi (\xi )=\arctan \left[ \Gamma
\tanh \left( \xi /2\xi _{0}\right) \right] $, $\xi _{0}=1/2(1-K^2)^{1/2}$, $%
\Gamma =[(1-K)/(1+K)]^{1/2}$, and $K=(\bar{\omega}-\Omega )/2\sqrt{v}$. These
solutions are parameterized by velocity $v$, frequency $\Omega $, phase
shift $\varphi $, and position $\tau $. The velocity $v$ is the amount by
which the wave is slowed from the copropagating frame velocity $u$.  Thus in 
the laboratory frame, the actual pulse velocity is $u-v$.  
The quantity $\xi _{0}$ must be
real, hence ${1-K^{2}}>0$. Thus the condition for existence of these
solutions is $|\bar{\omega}-\Omega |<2\sqrt{v}$. This choice of
parameters provides relatively simple mathematical expressions for the
solitary waves. In practice it is easier to both control and measure peak
amplitude, $A=2v^{3/4}(1-K)^{1/2},$ than the pulse velocity, therefore $A$, $%
\Omega $, $\varphi $, $\tau $ form a more suitable set of parameters. Given
the pulse amplitude $A$, the corresponding velocity parameter
depends on the value of the quantity $\bar{\omega}-\Omega $. If $\bar{\omega}%
=\Omega $, then $v=(A/2)^{4/3}$ trivially. For the case when $\bar{\omega}%
\not=\Omega $, write the amplitude as $A=2v^{3/4}(1-\sigma |\bar{\omega}%
-\Omega |/2\sqrt{v})^{1/2}$, where the parameter $\sigma =\mbox{sgn}(\bar{%
\omega}-\Omega )$. Then defining $\bar{v}=(2\sqrt{v}/|\bar{\omega}-\Omega
|-\sigma )^{1/2}$ and $\bar{A}=\sqrt{27/2}|\bar{\omega}-\Omega |^{-3/2}A$
leads to an expression for the rescaled velocity $\bar{v}=(y^{-1/3}-\sigma
y^{1/3})/\sqrt{3}$, where $y=\sigma \lbrack (\bar{A}^{2}+\sigma )^{1/2}-\bar{%
A}]^{1/2}$. In this calculation, the appropriate branches have been chosen
so that the expressions are consistent with reality and positivity
conditions on the parameters.

In optics it has become standard practice to refer to certain solutions of
nonintegrable systems as solitons. These solutions are characterized as
solitary waves which are robust to external perturbations including
collisions with other solitary waves. In addition, arbitrary initial data
for these ``soliton" supporting systems tends to evolve into a sum of
solitary waves and continuous
radiation. The remainder of this Letter details our numerical investigation
of these properties, in which Eqs.~(\ref{NormalizedEq}) are integrated in
the case of delta-distributed resonance frequencies and zero detuning.

Numerical simulations of the evolution of Gaussian initial data
are presented in Figs.~\ref{f1}a and b. In Fig.~\ref{f1}a the initial condition
$E(t,0)=\exp(-t^2/2)/2$ simply
evolves into continuous radiation, while in Fig.~\ref{f1}b the initial
condition $E(t,0)=2\exp(-t^2/2)$ emits some
radiation but also achieves energy confinement and persists as a soliton. This
behavior is similar to the self-induced transparency exhibited by the
Maxwell-Bloch equations~\cite{McCallHahn} which describe optical pulse
interaction with resonant two-level media. As the amplitude of the initial
pulse is increased, the pulse splits into two [illustrated in
Fig~\ref{f2}a for the initial condition $E(t,0)=5\exp(-t^2/2)$] or more
solitons and emits continuous
radiation. The inset shows a comparison between the numerics and
the analytic form of the solitary waves of Eq.~(\ref{solutions}). The velocity
and frequency of the solitary waves are obtained from measurements of their
amplitudes and half widths in our simulation.
The agreement of the analytic forms with the simulation results indicates that
the system self-selects the solitary waves presented in Eq.~(\ref{solutions}).
Fig.~\ref{f2}b shows the amplitude(s) of these solitons as a function
of input pulse amplitude. As the input pulse amplitude increases,
the output soliton amplitude also increases until a bifurcation occurs and
a new soliton emerges.  Increasing the input pulse amplitude further results
in the production of more solitons, along with continuous radiation.

Two simulations illustrating collision dynamics are presented in Figs.~\ref%
{f3}a and \ref{f3}b, where the sum of two well-separated solitary waves is
used as the initial condition. Fig.~\ref{f3}a illustrates an in-phase
collision, in which the relative phase $\Delta\varphi=\varphi_1-\varphi_2=0$, 
where the
subscripts identify the soliton. The other initial soliton
parameters are $v_{1}=1$, $\Omega _{1}=0$, $\tau _{1}=-10$%
, $v_{2}=2$, $\Omega _{2}=0$, and $\tau _{2}=0$. In Fig.~%
\ref{f3}b an out-of-phase $\Delta \varphi =\pi $ collision is illustrated.
The same parameters are used except for a shift in the relative phase.
This results in a much faster collision. In both simulations the
solitary waves persist after the interaction, although their
characteristic parameters undergo shifts and radiation is emitted during
the collision. A numerical study indicates that the
collisions are quasielastic for values in the approximate interval
$\Delta \varphi \in (\pi,2\pi)$. For some $\Delta\varphi$ values away from
this interval, simulations show that one of the solitons is completely
destroyed while the other persists.  A detailed analysis of the
dependence on initial parameters will be presented elsewhere.

The soliton phenomena described above occurs at light
intensities such that the dimensionless field amplitude $E$
is at least of order one. The intensity at which $E$ is order one
estimated as
$I\simeq(c/\varkappa )(m\omega_0^{3}/e)^{2}\simeq c(\hbar \omega_0 /ea)^{2}$.
For the particle radius $a=20$nm and the carrier wavelength $\lambda_0=500$nm,
this results in light intensity of $I\sim10$GW/cm$^2$ which can be easily
obtained with ultrashort laser pulses.  The optical pulse durations for which
this model is valid are limited by the condition $\Delta\omega\ll\omega_0$,
(the spectral width of the pulse must be much smaller than the carrier
frequency) required by the slowly varying envelope approximation.  The pulse
duration should also be much shorter than the characteristic plasmonic
oscillation damping time,
which is determined by the time required for electron thermalization in
the metal nanoparticles ($\sim$400fs~\cite{HGMPB,VCDFVPCLPB}).
The envelope approximation is appropriate for pulses with width
$\tau\gtrsim20$fs.

In summary, a family of solitary wave solutions is derived in the envelope
approximation for the Maxwell wave and Duffing oscillator equations, showing
that energy confinement is possible for resonant optical pulse
interaction with plasmonic oscillations in metal nanoparticles. The
existence condition for these solutions is presented. Numerical simulations
show that stable solitary waves evolve from arbitrarily-shaped initial
pulses with sufficient amplitudes and exhibit behavior analogous to
self-induced transparency in Maxwell-Bloch. Simulations also reveal that
the collision dynamics are highly dependent on initial soliton parameters,
behaving quasielastically in some regimes but having radically different
behavior in others. The authors are grateful to V. P. Drachev for helpful
discussions.  In addition, we would like to acknowledge funding under
Arizona Proposition 301, LANL, and NSF.

\section*{List of Figure Captions}

\begin{figure}[h]
\centerline{\scalebox{0.5}
  {\includegraphics{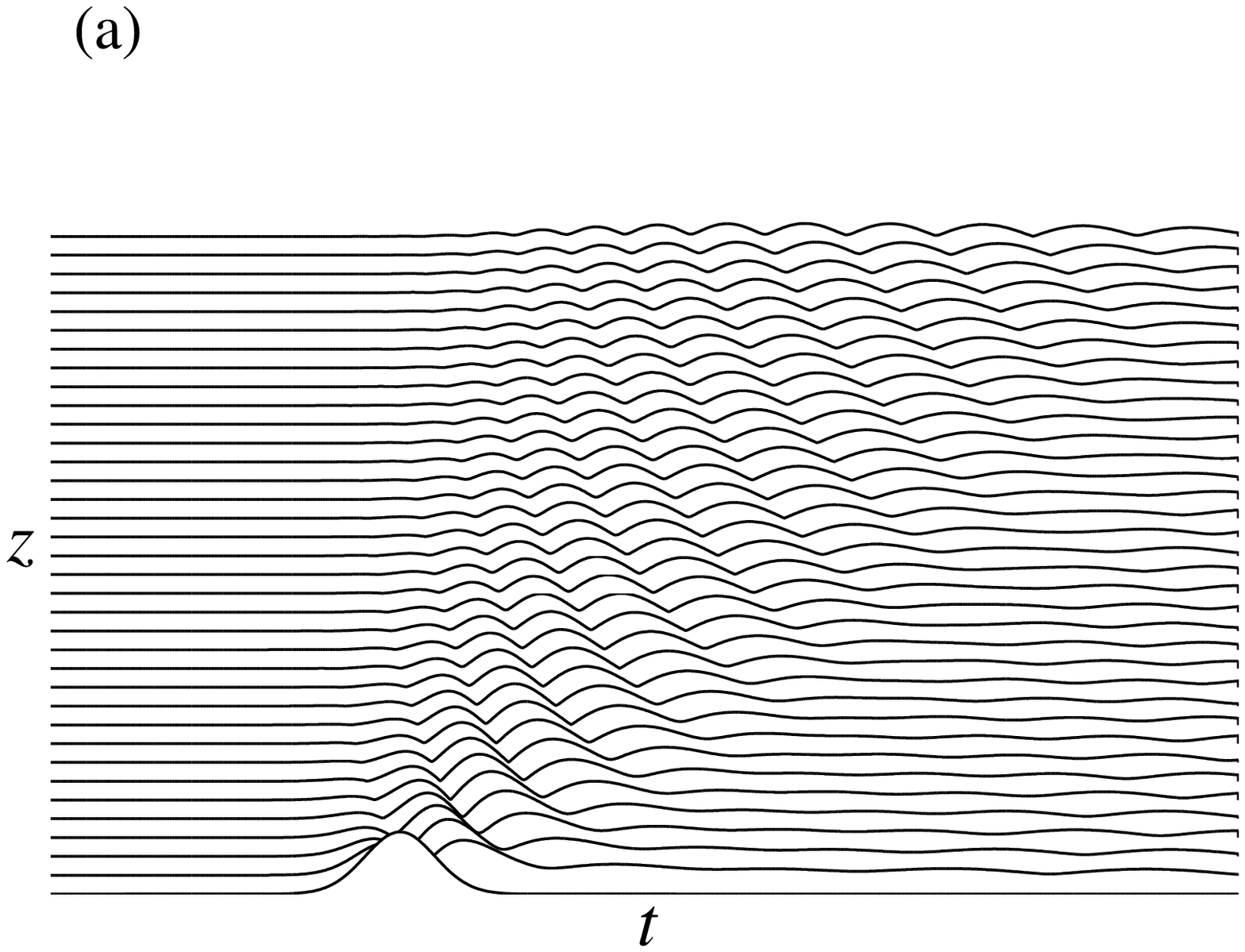}\includegraphics{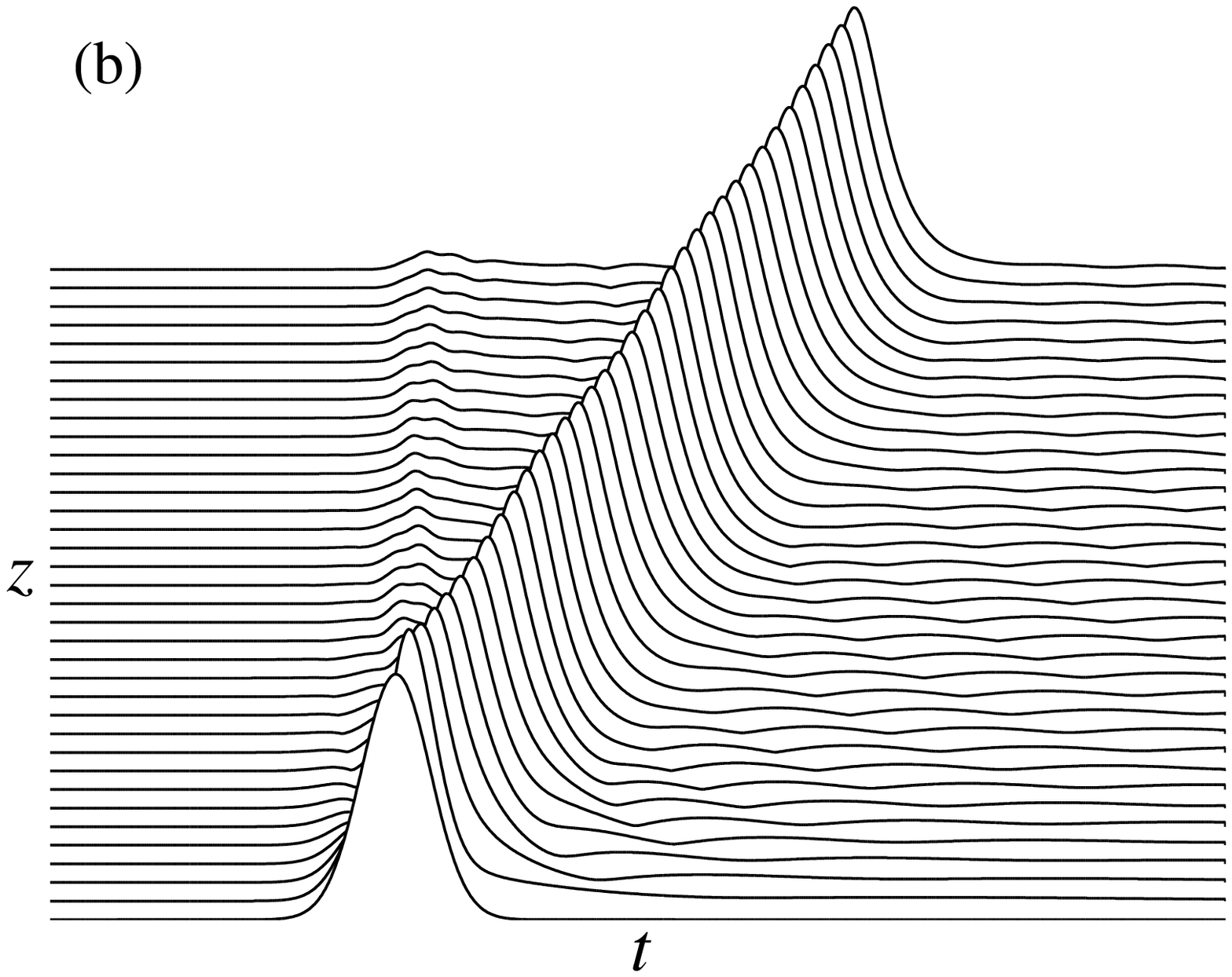}}}
\caption{Evolution of electric field amplitude with initial conditions $%
\exp(-t^2/2)/2$ (left) and $2\exp(-t^2/2)$ (right).}
\label{f1}
\end{figure}

\begin{figure}[h]
\centerline{\scalebox{0.34}{\includegraphics{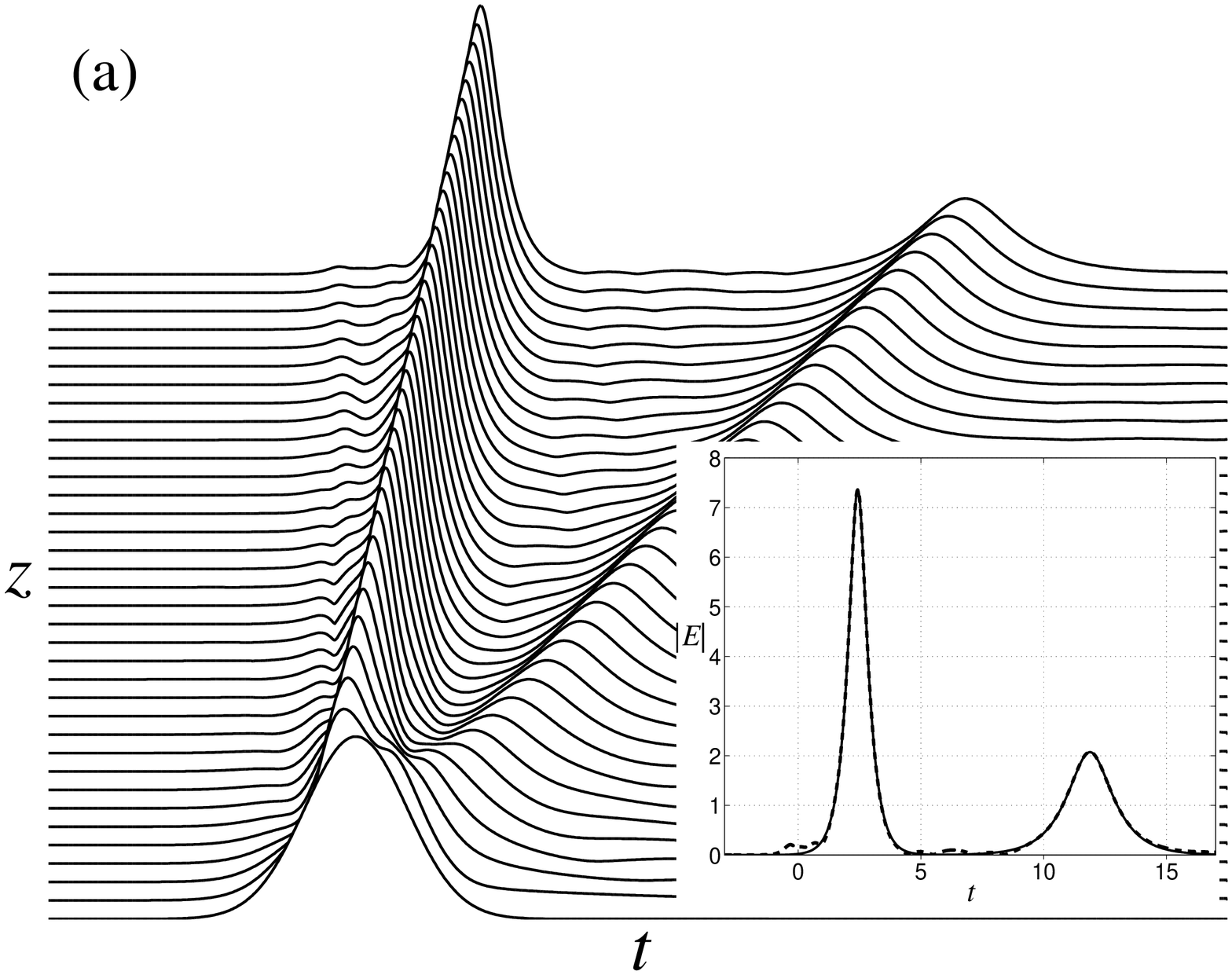}}
\scalebox{0.5}{\includegraphics{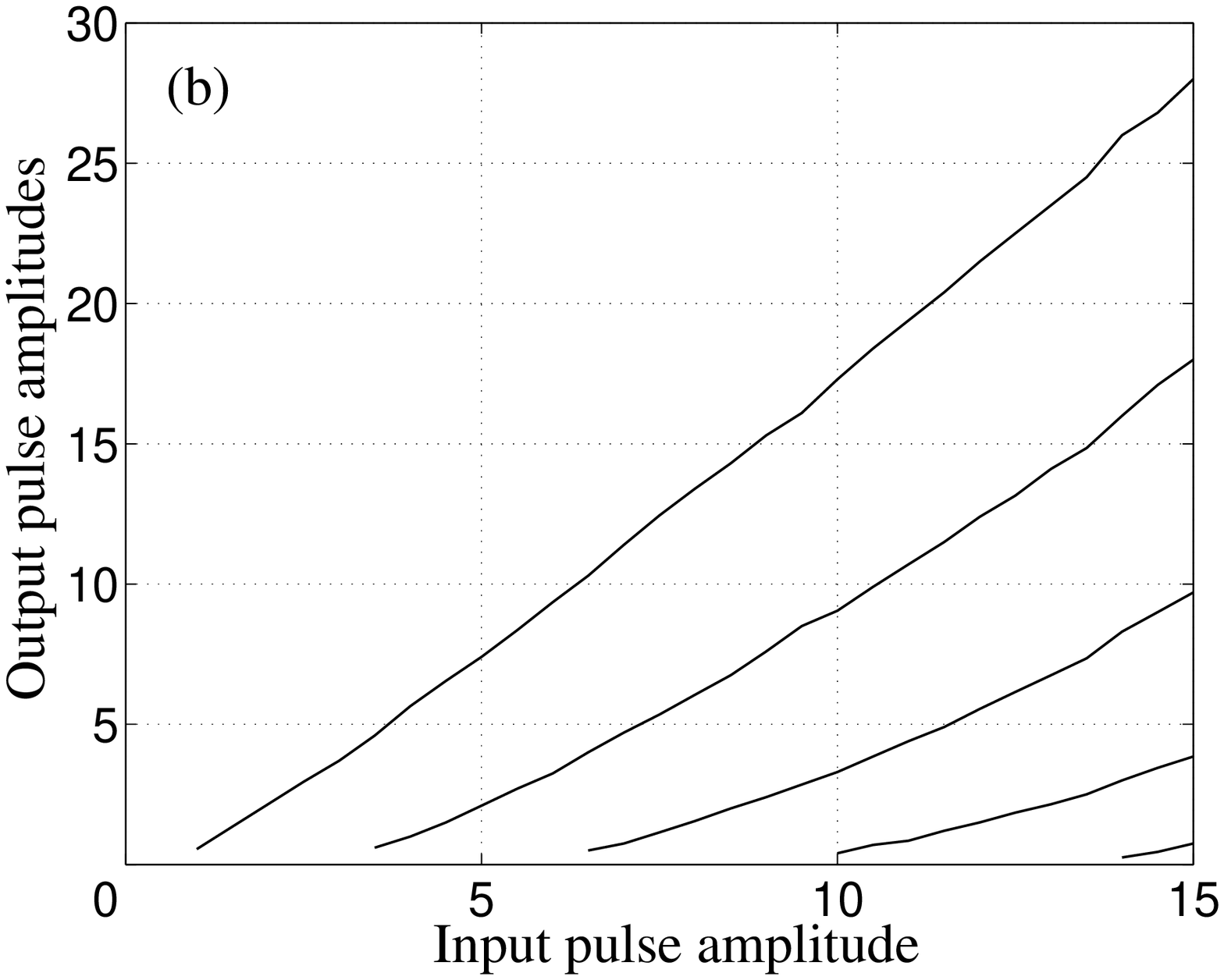}}}
\caption{Left: Evolution of electric field amplitude with initial condition
$5\exp(-t^2/2)$. The inset shows a comparison of the numerics (dashed line)
with the analytic form of the solitary wave solutions (solid line). Right:
output solitary wave amplitude(s) as a function of Gaussian input pulse
amplitude $A_0$, where the initial condition is given by $A_0\exp(-t^2/2)$.}
\label{f2}
\end{figure}

\begin{figure}[h]
\centerline{\scalebox{0.5}
  {\includegraphics{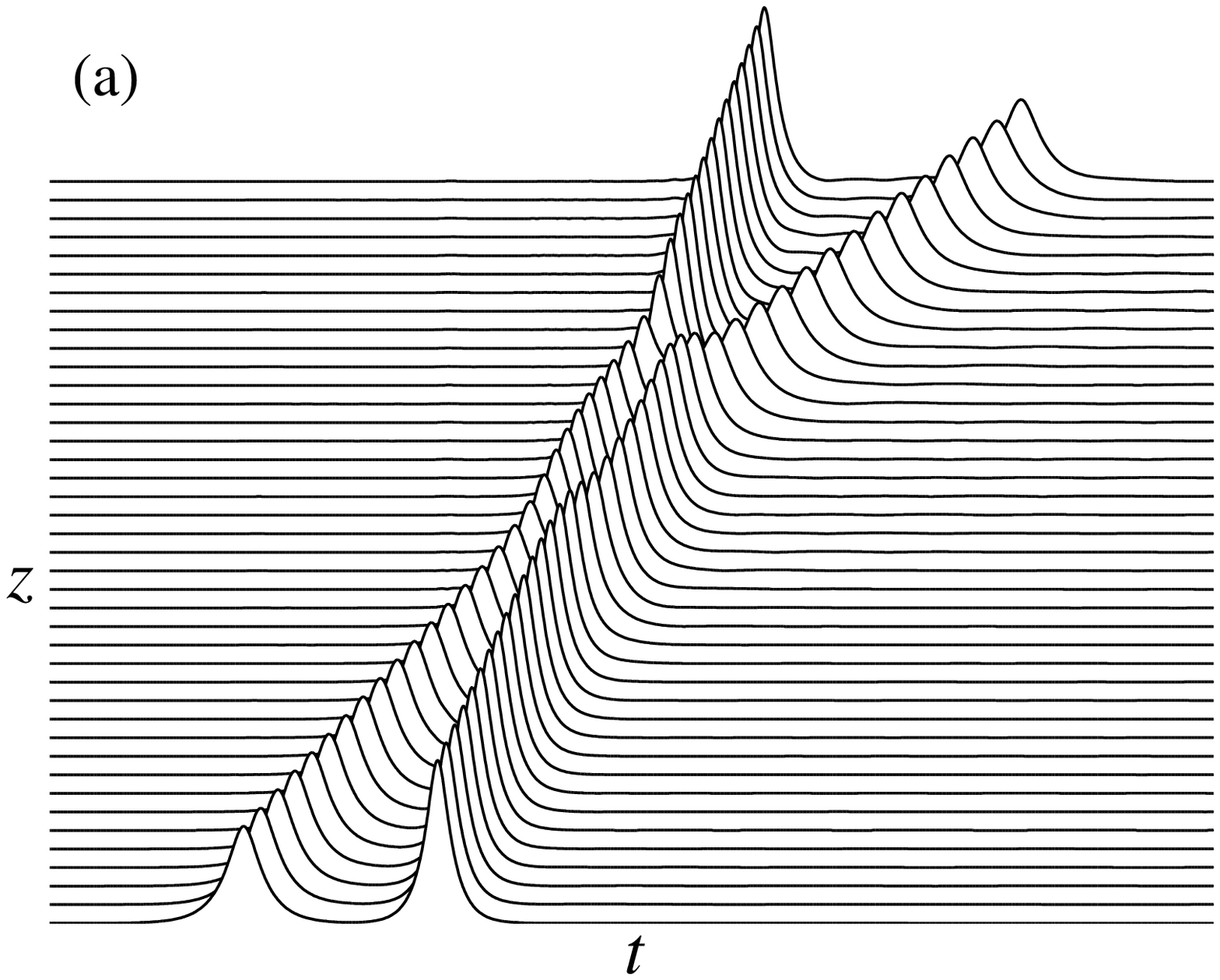}\includegraphics{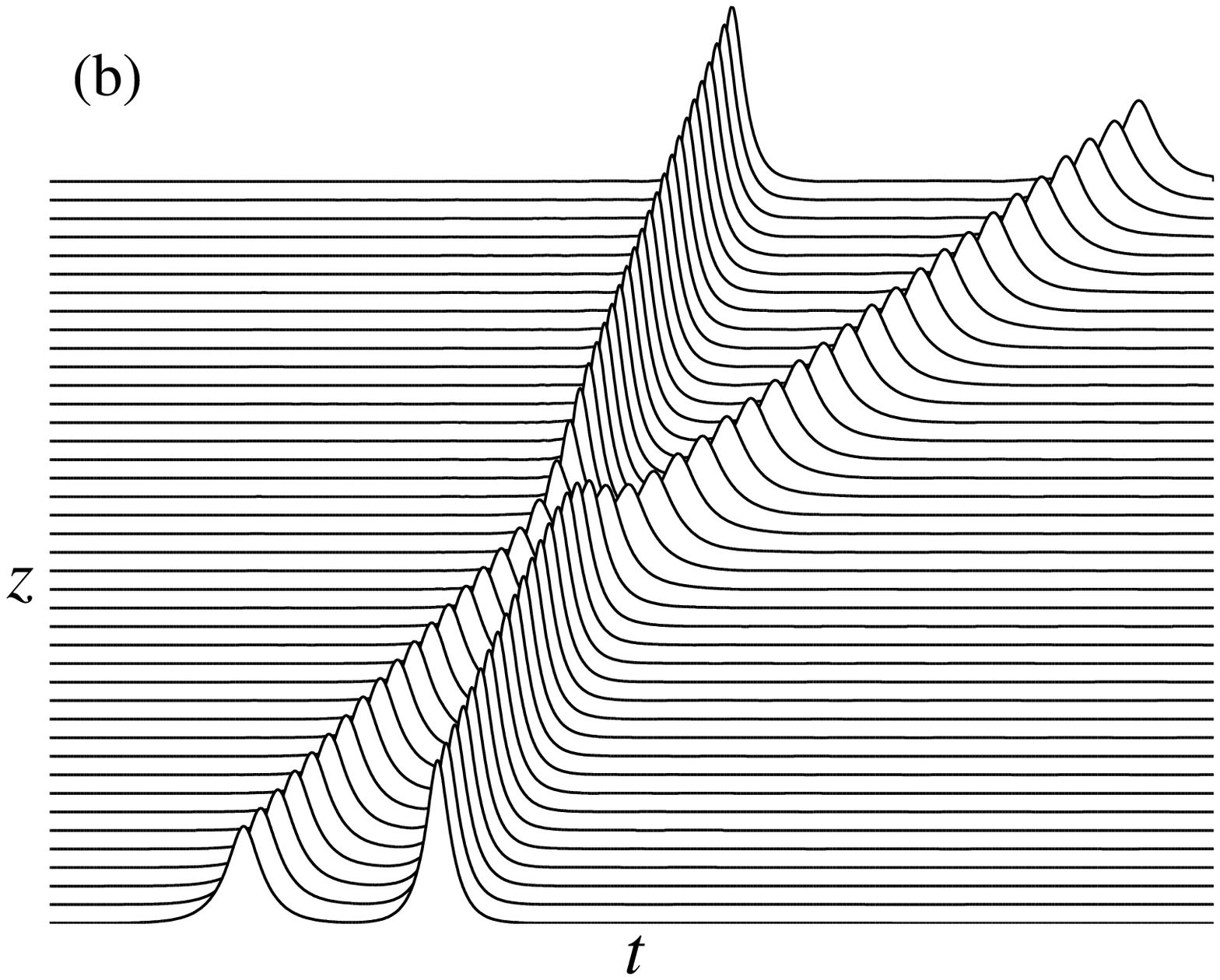}}}
\caption{Electric field amplitude showing collision dynamics of solitons for
different values of relative phase. Left: $\Delta\varphi=0$; right: $%
\Delta\varphi=\pi$.}
\label{f3}
\end{figure}

\ref{f1} Evolution of electric field amplitude with initial conditions $%
\exp(-t^2/2)/2$ (left) and $2\exp(-t^2/2)$ (right).\newline
\ref{f2} Left: Evolution of electric field amplitude with initial condition
$5\exp(-t^2/2)$. The inset shows a comparison of the numerics (dashed line)
with the analytic form of the solitary wave solutions (solid line). Right:
output solitary wave amplitude(s) as a function of Gaussian input pulse
amplitude $A_0$, where the initial condition is given by $A_0\exp(-t^2/2)$.
\newline
\ref{f3} Electric field amplitude showing collision dynamics of solitons for
different values of relative phase. Left: $\Delta\varphi=0$; right: $%
\Delta\varphi=\pi$. 



\end{document}